\documentclass[amsmath,amssymb,showpacs,floatfix,preprintnumbers,lengthcheck,prd,superscriptaddress,twocolumn]{revtex4}
\usepackage{graphicx}
\usepackage{multirow}

\newcommand{\beq}{\begin{equation}}
\newcommand{\eeq}{\end{equation}}
\newcommand{\beqs}{\begin{eqnarray}}
\newcommand{\eeqs}{\end{eqnarray}}

\begin{document}

\title{Lattice Simulations and Infrared Conformality}

\author{T.~Appelquist}
\affiliation{Department of Physics, Sloane Laboratory, Yale University,
             New Haven, Connecticut 06520, USA}

\author{G.~T.~Fleming}
\affiliation{Department of Physics, Sloane Laboratory, Yale University,
             New Haven, Connecticut 06520, USA}

\author{M.~F.~Lin}
\affiliation{Department of Physics, Sloane Laboratory, Yale University,
             New Haven, Connecticut 06520, USA}

\author{E.~T.~Neil}
\affiliation{Theoretical Physics Department, Fermi National Accelerator Laboratory, Batavia, Illinois 60510, USA}

\author{D.~A.~Schaich}
\affiliation{Department of Physics, Boston University, Boston, Massachusetts 02215, USA}

\begin{abstract}
We examine several recent lattice-simulation data sets, asking whether they are consistent with infrared conformality. We observe, in particular, that for an $SU(3)$ gauge theory with $12$ Dirac fermions in the fundamental representation, recent simulation data can be described assuming infrared conformality. Lattice simulations include a fermion mass $m$ which is then extrapolated to zero, and we note that this data can be fit by a small-$m$ expansion, allowing a controlled extrapolation. We also note that the conformal hypothesis does not work well for two theories that are known or expected to be confining and chirally broken, and that it does work well
for another theory expected to be infrared conformal.

\end{abstract}

\pacs{11.30.Qc,11.25.Hf,11.15.Ha,12.60.Nz~~~~~~~~~~~~~~~~~~FERMILAB-PUB-11-269-T}


\maketitle

\section{\textbf{Introduction}}

During the past few years, lattice simulations have been employed to study the infrared behavior of a variety of
gauge theories that could be relevant to physics beyond the standard model. Since much of the numerical code was originally developed for QCD, many of the simulations have focused on an $SU(3)$ gauge theory with varying numbers of massless fermions.

The lattice simulations of Refs. \cite{Appelquist:2007hu,Appelquist:2009ty} considered both $8$ and $12$ massless fermions in the fundamental representation, concluding that the former is a confining and chirally broken theory like QCD, and indicating that the latter is conformal in the infrared, dominated by a fixed point. Since then, various authors have examined the $12$-fermion theory, some agreeing that it is indeed infrared conformal \cite{Deuzeman:2009mh, Hasenfratz:2010fi, Itou:2010we, Hasenfratz:2011xn}, but others arguing that it is confining
and chirally broken \cite{Fodor:2011tu,Jin:2009mc}. The $SU(3)$ theory with $10$ massless Dirac fermions in the fundamental representation has also been studied, with the conclusion that it is infrared conformal \cite{Yamada:2010wd}.

The recent study of the $12$-fermion $SU(3)$ theory by Fodor et al \cite{Fodor:2011tu} is particularly interesting
because the simulation data set covers a wide range of fermion mass values, and finite-volume effects are relatively small. Their analysis leads them to the conclusion that their   simulation data for masses, the pseudoscalar decay constant and the chiral condensate are more compatible with confinement and chiral symmetry breaking in the massless limit.

In this paper, we examine the simulation data of Ref. \cite{Fodor:2011tu} noting that it can also be  described assuming that the massless theory is conformal in the infrared. Lattice data for several other theories are also considered. Since lattice simulations are carried out by including a fermion mass $m$ that is then extrapolated to zero, a question for any fit is whether the lattice data can be interpreted in terms of a small-$m$ expansion, allowing for a controlled extrapolation to zero. We argue that this is the case with the conformal hypothesis.  Finite-volume effects are considered and shown to be relatively small.

As a check on this conclusion, we attempt a similar fit to an $SU(3)$ theory with $2$ fermions in the fundamental representation, which is known to be in the broken phase, and an $SU(3)$ theory with $6$ fermions in the fundamental representation, which is strongly believed to be in the broken phase. In each case the quality of the fit is poor. We also examine the lattice data of Bursa et al \cite{Bursa:2011ru} for an $SU(2)$ gauge theory with $2$ fermions in the adjoint representation, which is believed to be conformal in the infrared \cite{Bursa:2009we, Catterall:2009sb, Hietanen:2009az, DeGrand:2011qd}. Like the $SU(3)$ theory with $12$ fermions, the data can be well fit by the conformal hypothesis, with a controlled extrapolation to $m = 0$.


\section{\textbf{The Conformal Framework}}

We first describe the scaling behavior we use to fit the lattice data of Ref. \cite{Fodor:2011tu}. The discussion is similar to that in Refs. \cite{DelDebbio:2010ze, DelDebbio:2010jy}, except that we also include nonleading terms in the scaling behavior. We assume that the infrared fixed point $g^{\star}$ approximately governs the behavior of the theory below some scale $\Lambda$, which, in a lattice setting, we take to be the inverse lattice spacing.  Finite-volume corrections will be described in Sec.~IV.

An explicit fermion mass, $m(\Lambda) \equiv m $ is introduced, with $m \ll \Lambda$. At scales below $\Lambda$, the running mass takes the form
\beq
m(\mu) = m ~( \Lambda / \mu)^ {\gamma^\star},
\eeq
where $\gamma^\star > 0$ is the mass anomalous dimension evaluated at the fixed point $g^{\star}$. At some scale $M \ll \Lambda$, the running mass satisfies
\beq
m(M) = M.
\eeq
At scales below $M$, the fermions decouple, and the running coupling flows away from the fixed point, triggering confinement. If the would-be fixed-point coupling $g^{\star}$ is reasonably strong, the induced confinement scale is of order $M$. We assume this to be the case.

The mass of each physical state $X$ is then set by the scale $M$. That is, using Eqs. 1 and 2, $ M_X \simeq C_{X}~ m^{[1 /( 1+ \gamma^{\star})]}$ \cite{Miransky:1998dh}, where the masses are expressed in units of $\Lambda$, and $C_X$ is a dimensionless coefficient not far above unity. In addition, there are correction terms, the largest of which in a small-$m$ expansion is of order $m$. Keeping only these two terms, we have
\beq
M_X = C_{X}~ m^{[1 /( 1+ \gamma^{\star})]} + D_{X}~  m.
\eeq
Since the explicit breaking of chiral symmetry is of order $M$ at the induced confinement scale $M$, there is no approximate chiral symmetry to be broken spontaneously. Thus this scaling law applies as well to the pseudoscalar state. The exponent $[1 /( 1+ \gamma^{\star})]$ is universal. 

Fodor et al \cite{Fodor:2011tu} also compute the pseudoscalar decay constant $F$ and the chiral condensate $\langle{\bar\psi} \psi\rangle$ as a function of $m$. Although $F$ plays no special role in the absence of spontaneous chiral symmetry breaking, we include it in our fit, using an expression similar to that for the masses:
\beq
F = C_{F} m^{[1 /( 1+ \gamma^{\star})]} + D_{F}  m .
\eeq

The chiral condensate, defined at the cutoff scale $\Lambda$, also vanishes as $m \rightarrow 0$. The leading, small-$m$ term is purely ultraviolet. This is the ``contact term," proportional to $m \Lambda^2$, independent of the form of the RG running of the coupling and $m(\mu)$. The second, coming from the RG running of $\langle{\bar\psi} \psi \rangle$ from $M$ to $\Lambda$, is proportional to $ M^{(3 - \gamma^{\star})}\Lambda^{\gamma^{\star}}$. Using Eqs. 1 and 2 to express $M$ in terms of $m$ and $\Lambda$, we have
\beq
\langle{\bar\psi} \psi \rangle = A_{C} m  + B_{C} m^{[(3 - \gamma^{\star}) /( 1+ \gamma^{\star})]} + .....   ,
\eeq
where now, as in Eqs. 3 and 4, all dimensionful quantities are expressed in terms of $\Lambda$, the inverse lattice spacing. The coefficients are dimensionless, and $m$ is the lattice mass.

In addition to these terms, we expect a contribution of order $M^3$, analogous to the leading-order terms in
$M_X$ and $F$, arising from the induced confinement scale $M$. And as with $M_X$ and $F$, there are further corrections, one of which is of order $m^3$. We therefore take
\begin{eqnarray}
\langle{\bar\psi} \psi \rangle = A_{C}m  +  B_{C} m^{[(3 - \gamma^{\star}) /( 1+ \gamma^{\star})]}  \nonumber \\
+ C_{C} m^{[3 /( 1+ \gamma^{\star})]}  + D_{C} m^3.
\end{eqnarray}
It will turn out that $0 < \gamma^{\star} < 1 $, so that these four terms also provide the basis for a small-$m$ expansion.


\section{\textbf{Fitting the Lattice Data Neglecting the $D$ terms}}

We fit the lattice data of Ref. \cite{Fodor:2011tu} for the masses of the scalar, pseudoscalar, vector, axial vector, nucleon, and parity partner of the nucleon, for the pseudoscalar decay constant, and for the condensate, first setting the $D$-term coefficients $D_X$, $D_F$, and $D_C$ to zero. We then ask whether the inclusion of the $D$ terms as well as finite-volume corrections improves the quality of the fit.

The simulations of Ref. \cite{Fodor:2011tu} were performed using a tree-level, Symanzik-improved gauge action,
 with lattice gauge coupling $\beta \equiv 6/g^2 = 2.2$. We assume here that this lattice coupling is consistent with the theory being approximately described by the infrared-fixed-point value of the running coupling throughout the range $M < \mu < \Lambda$.

  The simulations were done for fermion masses $m = 0.035$, $0.0325$, $0.030$, $0.0275$, $0.025$, $0.020$, $0.015$, $0.010$ (in lattice units), with lattice volume $24^{3} \times 48$ for the heaviest $4$ masses, with volume $32^{3} \times 64$ for $m = 0.025$, with volume $40^{3} \times 80$ for $m = 0.020$, and with three volumes ranging up to $48^{3} \times 96$ for $m = 0.015$ and $0.010$. In the fits reported here, we use the data at the largest volume available at each $m$ value \cite{Fodor:2011tu}.

\begin{figure}
\includegraphics[width=85mm]{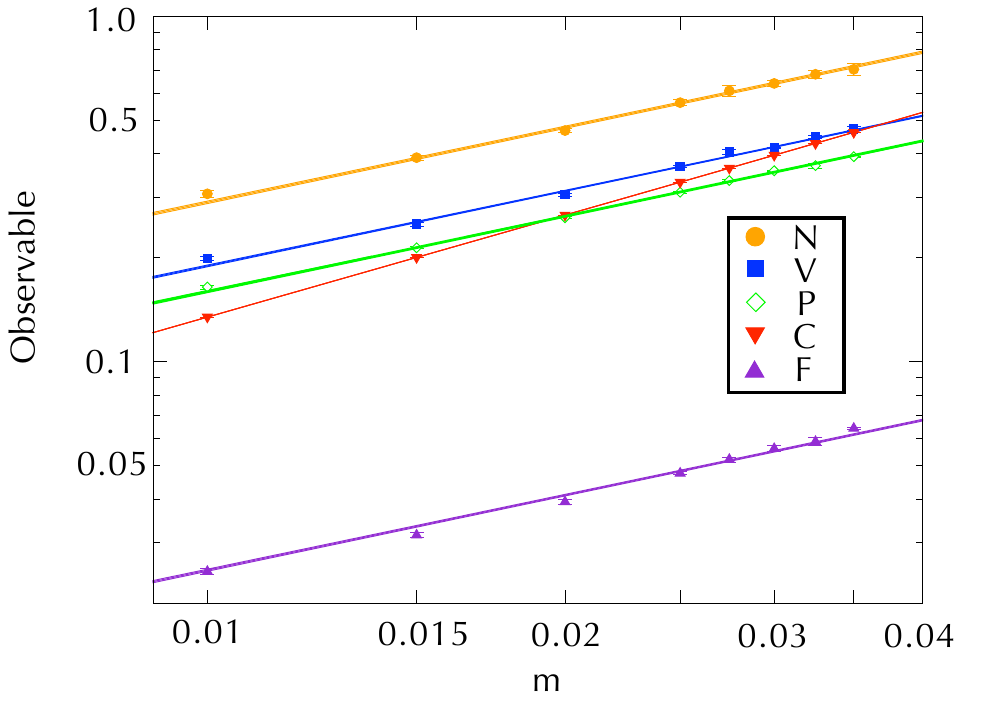}
\caption{ Log-log plot showing masses of the pseudoscalar (P), vector (V), and nucleon (N) states, the pseudoscalar decay constant (F), and the condensate (C) as a function of $m$, as reported in Ref. \cite{Fodor:2011tu}, along with our conformal fit to these quantities, with the $D$ terms set to zero. The universal slope of the P, V, N, and F curves provides a good fit to the simulation data.
\label{fig:combined} }
 \end{figure}

In Fig.~\ref{fig:combined}, we show the simulation data for the pseudoscalar (P), vector (V), and nucleon (N) masses, for the pseudoscalar decay condensate (F), and for the condensate (C) as a function of fermion mass $m$, along with our conformal fit to these quantities. The common log-log slope for P, V, N, and F, enforced by the universal scaling exponent $1/(1+\gamma^{\star})$, fits the data points well. The slope of the  condensate curve is determined dominantly by the leading, linear term of Eq. 6.

To further explore the conformal fit, we report on the left side of Table~\ref{tab:fit} the results of a fit to all the masses, the pseudoscalar decay constant, and the chiral condensate, with the $D$ terms set to zero and neglecting finite-volume effects. For this fit, the anomalous dimension is $\gamma^\star \approx 0.386 \pm 0.010$ and $\chi^2 / N = 2.508$, with $N = 53$ degrees of freedom.

Our (statistical) error analyses here and below do not take into account correlations of the numerical data, which would require access to the full simulation data set.  As a result, our errors may be underestimated and $\chi^2 / N$ may not directly indicate goodness-of-fit.  While this value of $\chi^2 / N$ is somewhat large, the fit reported so far is a very simple one. We have not yet included the $D$ terms of Eqs. 3, 4, and 6, and we have not taken into account possible finite-volume corrections. From the fit reported so far, the latter can be anticipated to be relatively small. The product $ML$ lies between $1.73$ and $2.23$, and finite-volume corrections should be small if $M_{X}L \propto ML \gg 1$. Inspection of the best-fit $C_X$ values in the left side of Table~\ref{tab:fit} indicates that this is the case.

In the next section, we will include finite-volume effects as well as 
the $D$ terms. The result, reported in the right-hand side of Table~\ref{tab:fit}, is quite 
encouraging. The finite-volume corrections, while not insignificant, are indeed 
relatively small, and the value of $\gamma^{\star}$ changes very little.  Furthermore, the fit improves, with $\chi^2/N = 0.944$ and $N=44$.

\begin{table}
\begin{tabular}{|c||c||c|c|c|}
\hline
\textbf{Obs.} & \mbox{\boldmath $D_X = 0$} & \multicolumn{3}{|c|}{\mbox{\boldmath $D_F \ne 0, z_X \ne 0$}} \\
\hline\hline
$\gamma^\star$ & 0.3858(98) & \multicolumn{3}{|c|}{0.403(13)} \\
\hline\hline
$C_P$ & 4.445(83) &  4.267(85) & $z_P$ & 0.209(64) \\
\hline
$C_S$ & 5.99(14) & 5.75(14) & $z_S$ & 0.63(45) \\
\hline
$C_V$ & 5.26(10) & 5.05(10) & $z_V$ & 0.319(88) \\
\hline
$C_A$ & 6.68(15) & 6.41(15) & $z_A$ & 0.50(30) \\
\hline
$C_N$ & 8.04(17) & 7.70(17) & $z_N$ & 0.35(18) \\
\hline
$C_{N^\star}$ & 8.06(17) & 7.73(17) & $z_{N^\star}$ & 0.49(24) \\
\hline\hline
$C_F$ & 0.692(13) & 0.455(39) & $z_F$ & 0.61(27)\\
$D_F$ & --- & 0.61(10) & &\\
\hline\hline
$A_C$ & 13.898(28) & 13.926(31) & $z_C$ & -0.036(43)\\
$B_C$ & -50.8(5.5) & -42.2(4.8) & & \\
$C_C$ & 94(11) & 79.0(9.5) & & \\
\hline\hline
$\chi^2 / \mathrm{dof}$ & $133 / 53$ & \multicolumn{3}{|c|}{$42 / 44$}\\
\hline
\end{tabular}
\caption{For the $SU(3)$ theory with $12$ fermions in the fundamental representation, best-fit results to the data of Ref. \cite{Fodor:2011tu}, for our global conformal fit as described in the text. On the left-hand side, all the $D$ terms are set to zero, and finite-volume corrections are neglected. On the right-hand side, $D_F$  [Eq. 4] is included, as well as finite-volume corrections, with $z_X$ as explained in Sec.~IV. The letters $S$, $P$, $V$, $A$, $N$, and $N^{\star}$ correspond respectively to the scalar, the pseudoscalar, the vector, the axial vector, the nucleon, and the parity partner of the nucleon. $F$ refers to the pseudoscalar decay constant and $C$ to the condensate. For each quantity, there are $8$ data points, one for each $m$ value.  The fits do not take into account possible correlations between different observables, as discussed in the text. \label{tab:fit}}
\end{table}

\begin{figure}
\includegraphics[width=85mm]{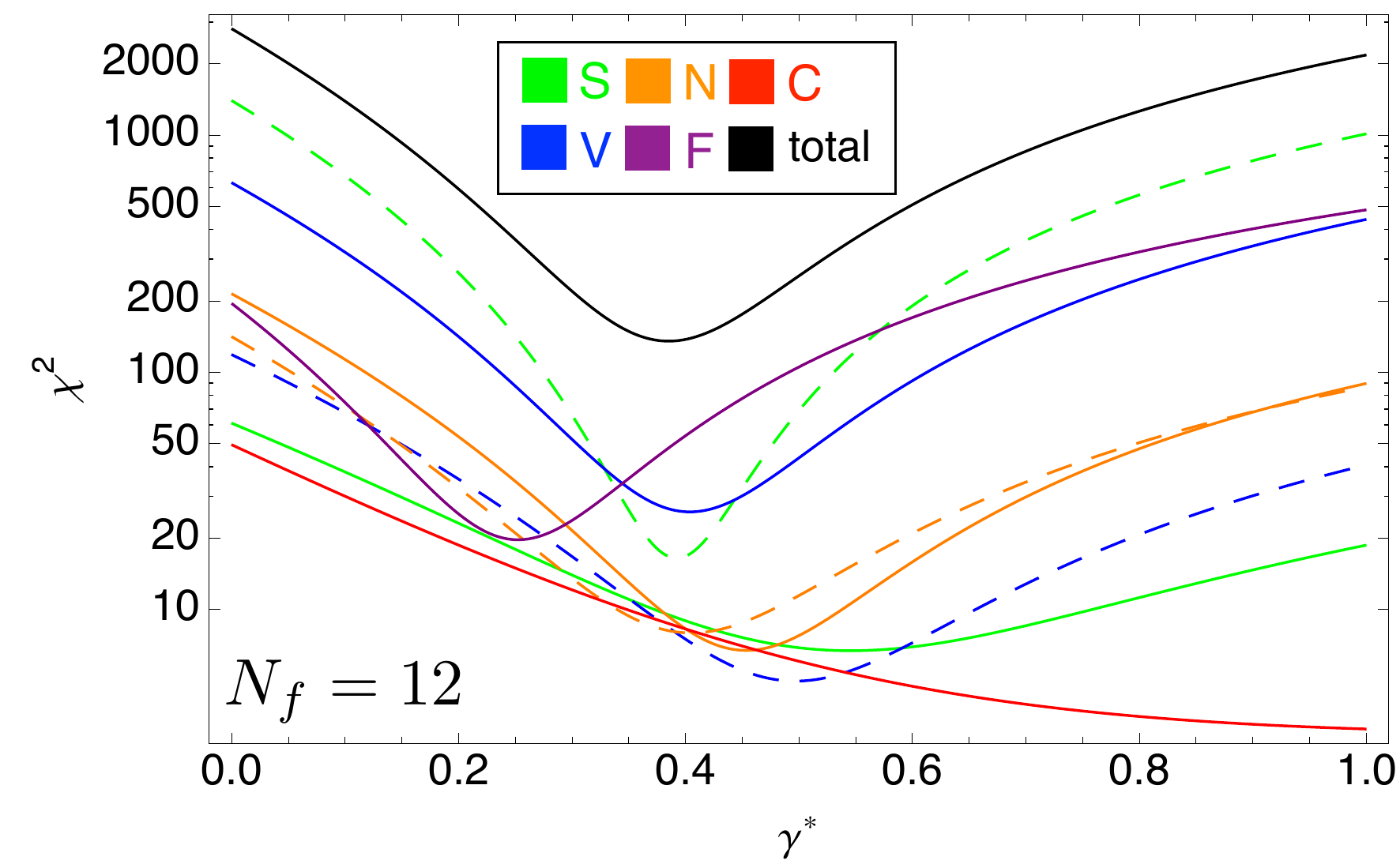}

\caption{For the $SU(3)$ theory with $12$ fermions in the fundamental representation, individual contributions to total $\chi^2$ from each channel, for a range of fixed values $0 < \gamma^\star < 1$.  Here, all $D$ terms are set to zero.  Fits are to the $N_f = 12$ data of Ref. \cite{Fodor:2011tu}.  Parity-odd states ($P, A, N^{\star}$) are shown as dashed curves in the same color as their parity partners. The black curve is the total $\chi^2$. \label{fig:gammascan12}}
\end{figure}

Before including the higher-order corrections, we examine the consistency of the conformal fit described above by performing separate fits to each mass, as well as to $F$ and the condensate. We fix a value of $\gamma^{\star}$ in the range $0 < \gamma^{\star} < 1$, and plot the $\chi^2$ for each fit as a function of $\gamma^{\star}$ in this range. The result is shown in Fig.~\ref{fig:gammascan12}, along with the sum of the individual $\chi^2$'s. The internal consistency is evident, with the minimum $\chi^2$ for each mass occurring at a similar value of $\gamma^{\star}$, and the minimum for the pseudoscalar decay constant at a value only slightly smaller. The condensate makes only a small contribution to the overall $\chi^2$. The minimum of the black curve corresponds to the total $\chi^2$ of the left fit of Table~\ref{tab:fit}.

\section{\textbf{Higher-Order Terms and Finite-Volume Effects}}

We next comment on the role of the $D$ terms in Eqs. 3, 4 and 6, 
inserting one $D$ term at a time and repeating the above global fit. At 
the same time, we consider the effect of finite-volume corrections.  We 
incorporate the latter by modifying Eqs. 3 and 4 to include first-order 
corrections in an expansion in $1/ML$, continuing to include the 
correction linear in the short-distance mass $m$. We thus take $M_X = 
C_{X}~ M~[ 1 + z_{X} / ML ] + D_{X}~ m$, with $M = m^{[1 /( 1+ \gamma^{\star})]}$, and similarly for $F$.  For the condensate, we simply 
replace $M = m^{[1 /( 1+ \gamma^{\star})]}$ in the second and third terms of Eq. 6  by $M~[ 1 + z_{C} / ML ]$.  While this modification is not claimed to be unique, it is physically sensible.  Furthermore, the finite-volume corrections are found to be insignificant for the condensate, so that the overall fit does not depend sensitively on the detailed form of this correction term.

For the data set of Ref. \cite{Fodor:2011tu}, the inclusion of the D 
term in any one of the masses, with or without the finite-volume 
corrections,  does not improve the quality of the fit. The value of each 
$D_X$ is consistent with $0$, with errors such that the $D_X$ [next-to-leading order (NLO)]
term is small compared to the $C_X$ [leading order (LO)] term for the full range of $m$ 
values. The inclusion of the $D_C$ (NNNLO) term in the condensate also 
does not improve the quality of the fit, with $D_C$ consistent with $0$ 
and the errors such that this term is relatively small throughout the 
range of $m$ values.

The inclusion of the $D_F$ (NLO) term in the pseudoscalar decay constant 
\emph{does} improve the global fit, as does the inclusion of the finite-volume correction terms $z_X$. In the right-hand side of Table~\ref{tab:fit}, we report the results of a fit including $D_F$ and all $z_X$.  The resulting anomalous dimension is $\gamma^\star = 0.403(13)$.  The values of $\gamma^{\star}$ and the $A$, $B$, and $C$ coefficients change very little from the left-hand fit.  We find $D_F = 0.61 \pm 0.10$, showing that the NLO term is small compared to the LO term for the full range of $m$ values.  The values of the $z_X$, $z_F$, and $z_C$ coefficients
show that finite-volume corrections are relatively small for the entire 
range of $m$ values, and extremely small for the condensate. Most
notably, we find $\chi^2 / N = 0.944$ with $N = 44$ degrees of freedom, again neglecting possible correlations between the observables.

In order to make a direct comparison to the broken-symmetry fit of Ref. \cite{Fodor:2011tu},
we have also applied our conformal fit to just the set of four quantities considered in their
global analysis: the pseudoscalar mass and decay constant, the nucleon mass, and the chiral
condensate. We find a best-fit value $\gamma^{\star} = 0.414 \pm 0.016$, and with other 
parameters in the same range as reported in Table~\ref{tab:fit}. Most importantly, we 
find $\chi^2 / N = 1.10$ with N = 20, to be compared to their broken-symmetry value of
$\chi^2 / N = 1.22$ with $N = 26$.

If a global fit including all the $D$ terms is attempted for the data set of Ref. \cite{Fodor:2011tu}, the $\chi^2$ dependence on $\gamma^{\star}$ becomes very flat, slightly favoring larger values in the range $0 < \gamma^{\star} < 1$, and leading to poor determinations of the D coefficients, with errors comparable in magnitude to the central values.  We conclude that the current data set is not extensive enough to perform a global fit with all the parameters of Eqs. 3, 4 and 6.  The availability of additional simulation data for larger $m$ values would be especially helpful in allowing a global fit that constrains the $D$ terms.

For the condensate, the constants $A_{C}$, $B_{C}$ and $C_{C}$ are such that the LO ($A_C$) term strongly dominates throughout the mass range. The NLO ($B_C$) term, also arising at scales of order $\Lambda$, and the NNLO ($C_C$) term, arising at scales of order $M$, are of opposite sign and strongly negatively correlated, with the NLO term dominating the NNLO term except for the largest, $m = 0.035$ point. 

That the fit leads to an NLO coefficient $B_C$ of opposite sign to the LO term is perhaps surprising, but we know of no reason why this correction cannot be of opposite sign. It is also possible that this is a consequence of the limited amount of simulation data for the condensate, which is strongly dominated by the LO, linear term. Furthermore, since the condensate is so sensitive to physics at the scale $\Lambda$, our assumption that the coupling can be approximated by its infrared fixed-point value out to $\Lambda$, which determines the form of the $B_C$ term, should break down first here.

Finally, it is interesting to note that for the masses, the $C_X$ coefficients of Table~\ref{tab:fit} are such that for small $m$, none of the states can decay into a combination of the others. Recalling that there is induced confinement in this theory at scale $M \propto m^{[1 /( 1+ \gamma^{\star})]}$, decay into the fundamental fermion and gauge-boson constituents is forbidden. Since there is no reason for other states such as the $0^{++}$ to be lighter, it appears that each of these states is stable for arbitrarily small $m$, but with an induced confinement radius diverging as $m \rightarrow 0$.

\begin{figure}
\includegraphics[width=85mm]{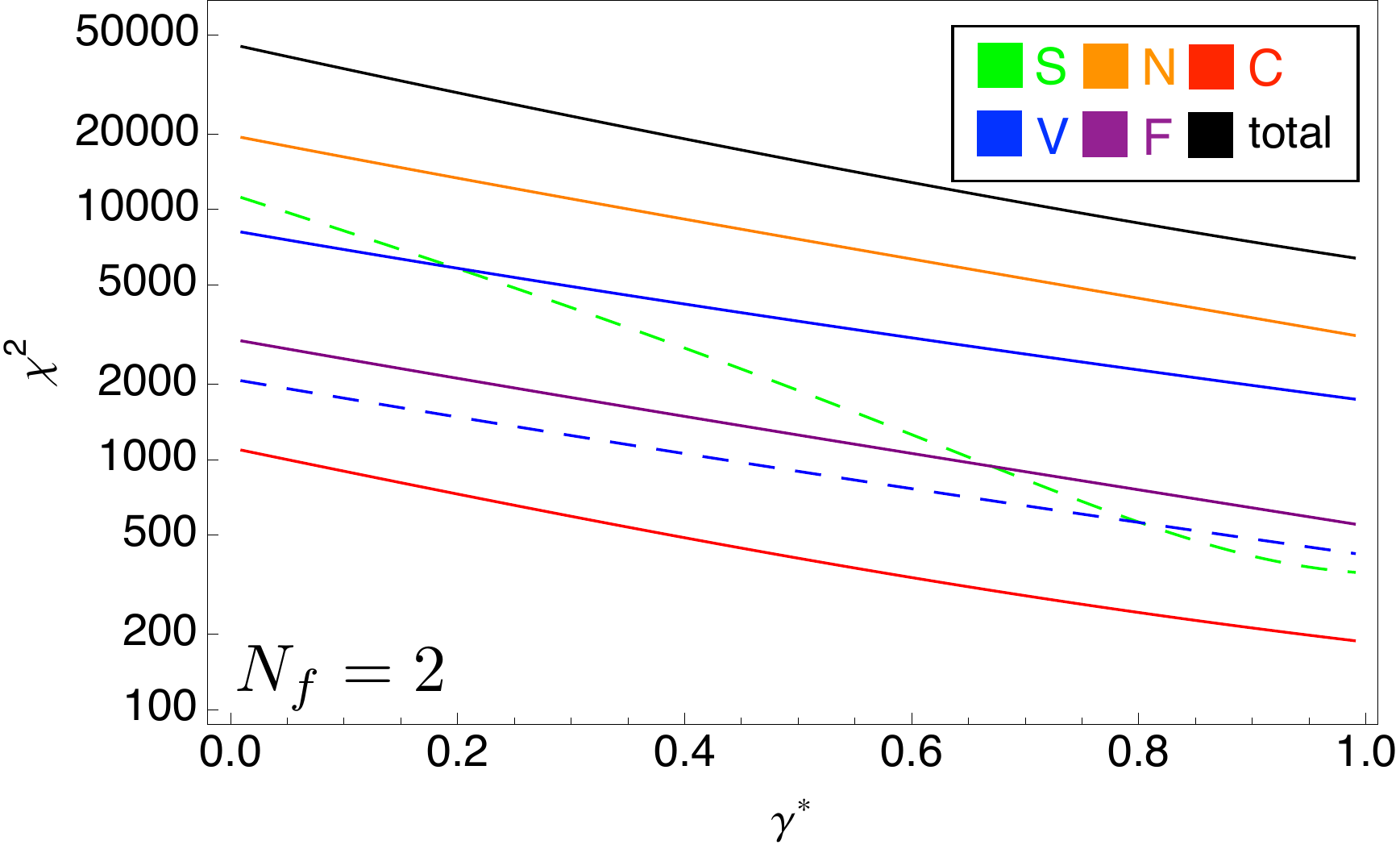}
\includegraphics[width=85mm]{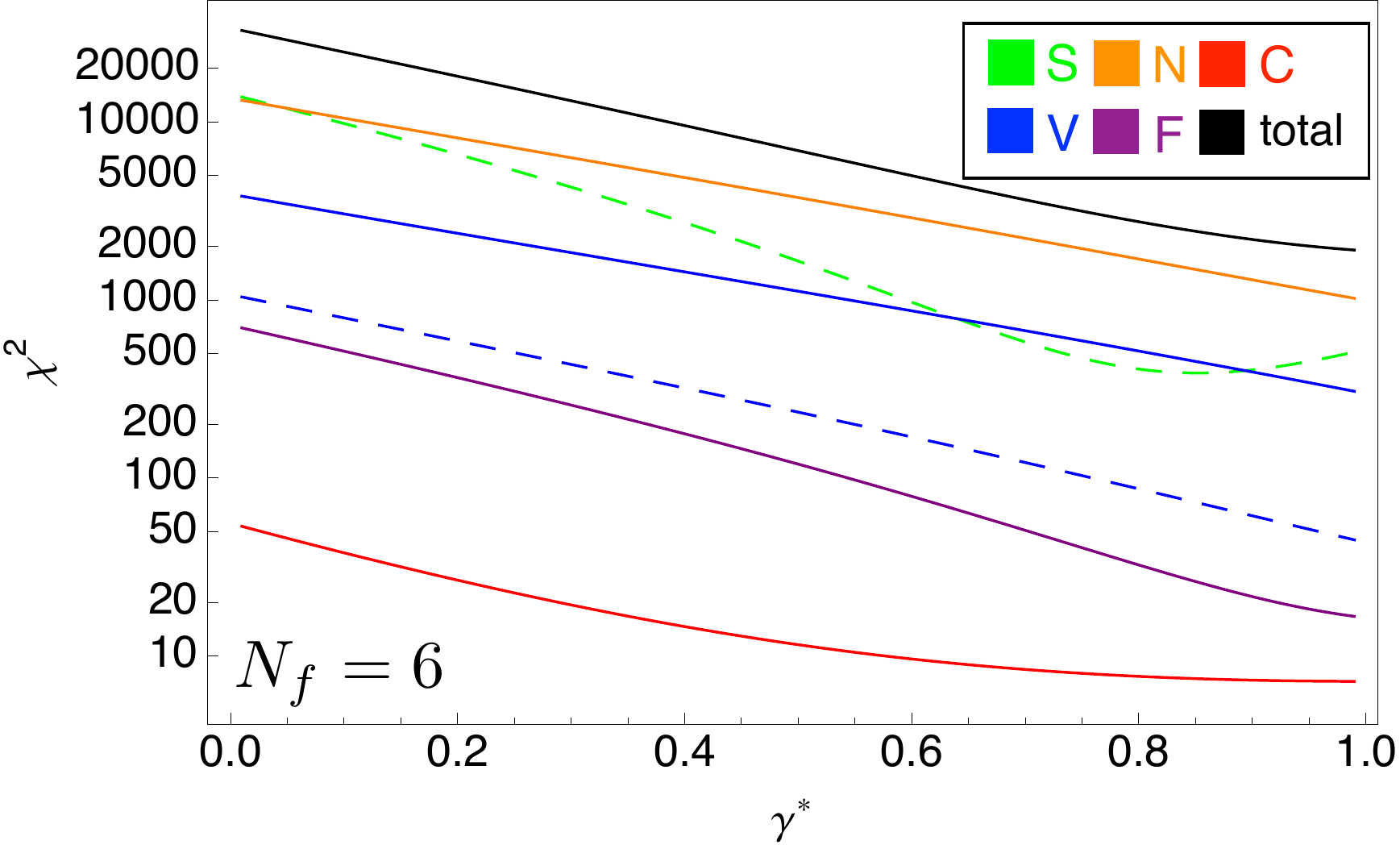}
\caption{For the $SU(3)$ theory with $2$ and $6$ fermions in the fundamental representation, individual contributions to total $\chi^2$ from each channel, for a range of fixed values $0 < \gamma^\star < 1$.  Total number of degrees of freedom is 27 in both cases.  All $D$ terms are set to zero.  The top (bottom) panel shows results from fitting to $N_f = 2$ ($N_f = 6$)  data, obtained from the simulations detailed in \cite{Appelquist:2009ka,Appelquist:2010xv}. Parity-odd states ($P, A$) are shown as dashed curves in the same color as their parity partners. \label{fig:gammascan26}}
\end{figure}

\section{\textbf{Testing Infrared Conformality on Other Gauge Theories}}

As a check on our conclusion that the simulation data of Ref. \cite{Fodor:2011tu} are consistent with infrared conformality, we have tested our conformal fit on two theories for which it should not work well and one for which it should. The former are an $SU(3)$ gauge theory with $2$ fermions in the fundamental representation, known to be in the broken phase, and an $SU(3)$ gauge theory with $6$ fermions in the fundamental representation, strongly believed to be in the broken phase. Here we fit the simulation data of  Refs.~\cite{Appelquist:2009ka, Appelquist:2010xv}.  In each case the quality of the fit is indeed poor, as shown in Fig.~\ref{fig:gammascan26}, plotting the $\chi^2$ for each individual fit, as well as the overall $\chi^2$, as function of $\gamma^{\star}$.  No clear minimum in $\chi^2$ appears for any channel except the $N_f = 6$ pseudoscalar mass, where it is at $\gamma^{\star}$ close to $1$. For $N_f = 2$, a minimum appears at $\gamma^{\star} \approx 1$. 
With chiral symmetry breaking, $M_P \sim m^{1/2}$ in lowest order chiral perturbation theory, corresponding effectively to $\gamma^\star = 1$.

It is worth noting that for the $SU(3)$ gauge theory with $2$ or $6$
fermions in the fundamental representation, the poorness of the conformal
fit should not be due to finite-volume effects.  Within the conformal
hypothesis, as we noted in the case of the $12$-fermion theory, a measure of
finite-volume effects is given by $ML = m^{1/(1+\gamma^{\star}) }L$. Here, the value of
$\gamma^{\star}$ emerging from the poor fit is of order unity, and $L = 32$, so $ML>  2.3$ for the entire range of $m$ values. Each of the associated masses is larger than $M$, so
finite-volume effects should be relatively small.

 We also note that for the $SU(3)$ gauge theory with $2$ fermions in the fundamental representation, a fit using chiral perturbation theory for a confining and chiral-breaking theory does work well \cite {Neil:2010sc}. For the $SU(3)$ gauge theory with $6$ fermions in the fundamental representation, a smaller set of fermion masses will be required to apply chiral perturbation theory \cite {Neil:2010sc}. But there is strong evidence from lattice simulations of the running coupling that this theory \emph{is} in the broken phase \cite{Appelquist:2007hu,Appelquist:2009ty}.

\begin{figure}
\includegraphics[width=85mm]{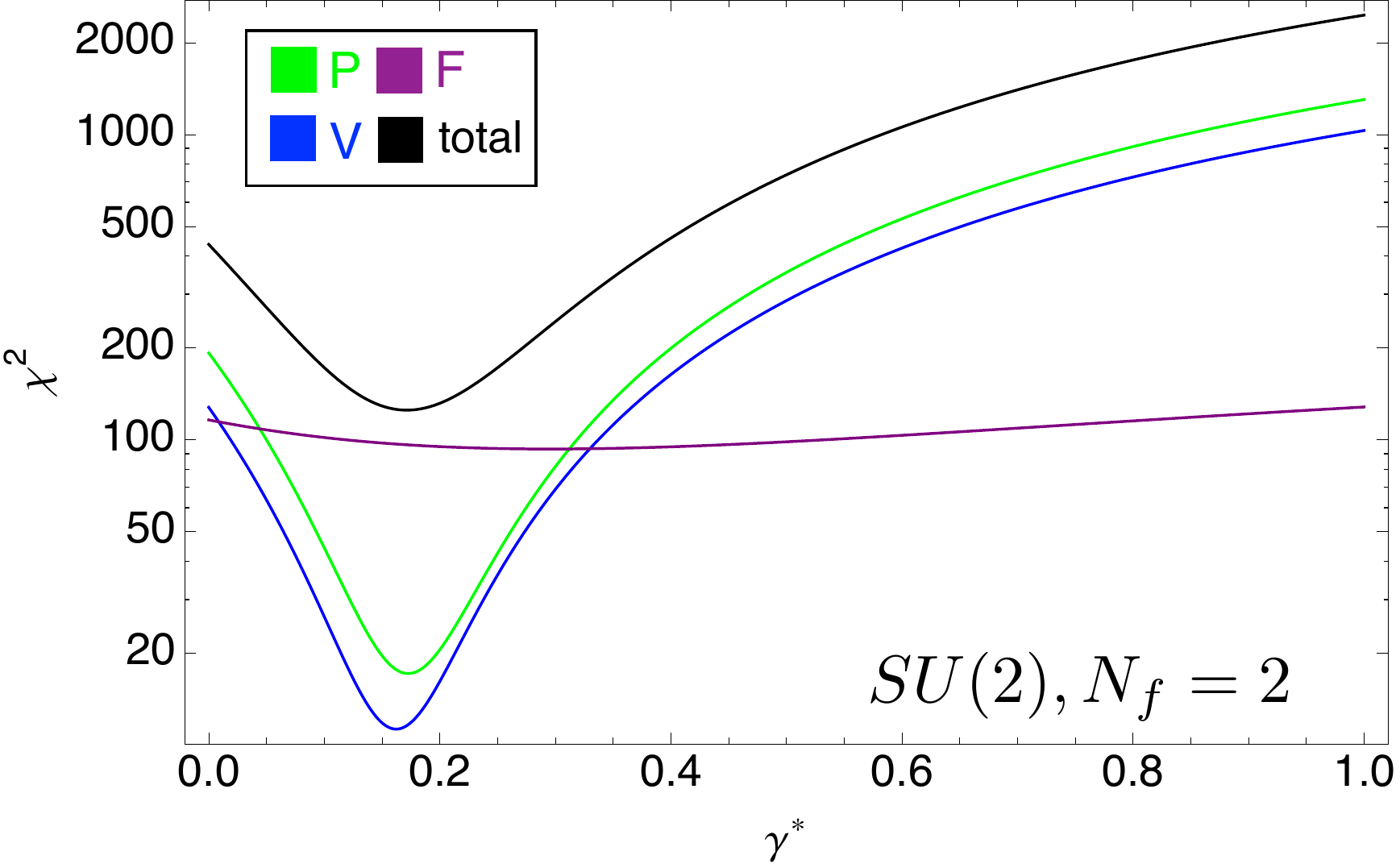}

\caption{For the $SU(2)$ theory with two fermions in the adjoint representation, individual contributions to total $\chi^2$ from each channel, for a range of fixed values $0 < \gamma^\star < 1$.  Total number of degrees of freedom is 14 (6 data points per channel).  All $D$ terms are fixed to zero.  Fits are to the SU$(2)$ two-flavor adjoint data of Bursa et al \cite{Bursa:2011ru}, with the restriction $m < 0.2$.   \label{fig:gammascan2c}}
\end{figure}

A theory for which a conformal fit \emph{should} work well is an $SU(2)$ gauge theory with $2$ fermions in the adjoint representation, widely believed to be conformal in the infrared \cite{Bursa:2009we, Catterall:2009sb, Hietanen:2009az,DeGrand:2011qd}. We have fit the simulation data of Bursa et al \cite{Bursa:2011ru} for the pseudoscalar and vector masses and pseudoscalar decay constant, assuming as before that $M$ is the induced confinement scale up to a coefficient of order unity. The data are used only in the range $m < 0.2$, in order to ensure that our formulas based on a small-$m$ expansion can be applied.  Figure \ref{fig:gammascan2c} shows $\chi^2$ versus $\gamma^\star$ for each channel, as well as the overall $\chi^2$, based on conformal fits as in Eqs.~3 and 4.  We find a clear minimum in $\chi^2$ at the best-fit value of $\gamma^\star = 0.17 \pm 0.05$.  This value is roughly consistent with previous determinations of $\gamma^\star$ \cite{Bursa:2009we,DeGrand:2011qd}.  Again, our analysis does not include the full data covariance matrix, so our error may be underestimated and $\chi^2 / N$ may not indicate goodness-of-fit.  The relatively large contribution of the decay constant to the overall $\chi^2$ may be due to underestimation of statistical errors, as the data as shown in Ref. \cite{Bursa:2011ru} are difficult to describe with any smooth function of $m$.  Addition of finite-volume corrections as described in Sec.~IV supports this conclusion, failing to improve the quality of the decay constant fit (but improving fits to the masses.)

\section{\textbf{Concluding Comments}}

We have argued that the simulation data of Ref. \cite{Fodor:2011tu} are consistent with the hypothesis that an $SU(3)$ gauge theory with $12$ massless fermions in the fundamental representation is conformal in the infrared. This conclusion is based on a simple fit to the data, in particular assuming that the gauge coupling can be approximated by its infrared-fixed-point value $g^{\star}$ out to the UV cutoff $\Lambda$ (the inverse lattice spacing). The mass anomalous dimension $\gamma$ is then set to its fixed-point value $\gamma^{\star}$.

A global fit including finite-volume and higher-order corrections yields $\chi^2 / N$ = 0.944 with $N = 44$.  We have argued that the fit describes a small-$m$ expansion 
covering the range of fermion masses used in the simulations, allowing a controlled extrapolation to $m = 0$.  It leads to a mass anomalous dimension  of $\gamma^{\star} = 0.403(13)$.  To compare directly with the broken-symmetry fit of Ref.~\cite{Fodor:2011tu}, we also fit to a  subset of four channels, finding $\chi^2 / N = 1.10$ with $N = 20$ as compared to the reported broken-symmetry value of $\chi^2 / N = 1.22$ with $N = 26$.

 Although not described in detail here, we have also used the infrared-conformal hypothesis to fit the simulation data of Ref. \cite{Fodor:2011tu} for the static quark potential. Since confinement is induced at scale $M$, an effective string tension $\sigma \propto M^2 \propto m^{[2/( 1+ \gamma^{\star})]}$ can be determined from the data assuming that string breaking has not yet set in. The fit works well, with a value of $\gamma^{\star}$ in good agreement with the other fits and an acceptable $\chi^2$.
 
We stress that we have not argued conclusively that the simulation 
data of Ref. \cite{Fodor:2011tu} demonstrates that the $SU(3)$ theory with 12 massless fermions is infrared conformal.  The simulation data can be described with similar fit 
quality by the chirally broken functional forms used in \cite{Fodor:2011tu}, with a slope and 
intercept. But the large value of the slope term compared to the intercept, for the existing range of $m$  values, does not provide the basis for a small-$m$ expansion in the spirit of chiral 
perturbation theory with a controlled extrapolation to $m = 0$. Further simulations at 
additional $m$ values will help to distinguish the two scenarios.

As a check on the validity of our infrared-conformal fit to the $SU(3)$ theory with $12$ fermions, we have observed that it does not work well for an $SU(3)$ gauge theory with $2$ fermions in the fundamental representation, which is known to be in the broken phase, or an $SU(3)$ gauge theory with $6$ fermions in the fundamental representation, which is strongly believed to be in the broken phase. On the other hand, it does work well for an $SU(2)$ gauge theory with $2$ fermions in the adjoint representation, believed to be in the infrared-conformal phase. We have fit the pseudoscalar and vector masses and decay constants, with a $\gamma^{\star}$ consistent with other references. Here, too, we have fit the string tension induced at finite $m$ with an anomalous dimension in agreement with the other fits.

It will next be important to address several of the simplifying assumptions made in these fits. Finite-volume effects should be examined further, as well as corrections due to the running of the coupling and mass anomalous dimension at higher mass scales. Also, a possible hierarchy between $M$ and the induced confinement scale should be considered.

\section{\textbf{Acknowledgments}}
We thank the members of the LSD Collaboration (R.~Babich, M.~I.~Buchoff, R.~C.~Brower, M.~Cheng, M.~A.~Clark, S.~D.~Cohen, J.~Kiskis, J.~C.~Osborn, C.~Rebbi, G.~Voronov, and P.~Vranas) for many helpful comments.  We also thank Tom DeGrand and Biagio Lucini for useful exchanges, and one of us (TA) thanks L. Del Debbio for a very helpful discussion. This work was supported partially by DOE Grant Nos.~DE-FG02-92ER-4074 (TA) and No.~DE-FG02-91ER40676 (DS), and by NSF Grant No.~PHY-0801068 (GF and ML).  Fermilab is operated by Fermi Research Alliance, LLC under Contract No.~DE-AC02-07CH11359 with the U.S. Department of Energy.

\bibliography{ConformalFit-v3}

\end{document}